\documentstyle[epic]{article}

\begin{document}

\title{Probing the anisotropic velocity of light in a gravitational field: another
test of general relativity}
\author{Vesselin Petkov \\
Physics Department, Concordia University\\
1455~De~Maisonneuve~Boulevard West\\
Montreal,~Quebec,~Canada~H3G~1M8\\
E-mail:~vpetkov@alcor.concordia.ca}
\date{31 December 2000}
\maketitle

\begin{abstract}
A corollary of general relativity that the average velocity of
light between two points in a gravitational field is anisotropic
has been overlooked. It is shown that this anisotropy can be
probed by an experiment which constitutes another test of general
relativity.

\medskip

\noindent PACS numbers: 04.80.Cc, 04.20.Cv
\end{abstract}

Although the bending of light and the zero velocity of light
which tries to leave a gravitationally collapsing body are
indications that the propagation of light in a gravitational
field is anisotropic, that anisotropy has been barely mentioned
in the case of the coordinate velocity of light only \cite
{schiff}, \cite{ohanian}. It has been overlooked that the average
velocities of light toward and away from a gravitational center
between two points of different gravitational potential are not
equal to $c$ and not the same. As we shall see bellow the
propagation of light in non-inertial reference frames
(accelerating or at rest in a gravitational field) is anisotropic
(direction-dependent). This can most clearly be demonstrated by
revisiting the issue of propagation of light in the Einstein
elevator experiment \cite{infeld} and studying the propagation of
light rays parallel to the elevator's acceleration (in addition
to the horizontal ray originally considered by Einstein).

Consider first an elevator accelerating with an acceleration $a=\left| {\bf a}%
\right| $ which represents a non-inertial (accelerating) reference frame $%
N^{a}$ (Figure 1). Three light rays are emitted simultaneously in
the elevator (in $N^{a}$) from points $D$, $A$, and $C$ toward
point $B$. Let $I$ be an inertial reference frame instantaneously
at rest with respect to $N^{a} $ (i.e. the comoving frame) at the
moment the light rays are emitted. The emission of the rays is
therefore simultaneous in $N^{a}$ as well as in $I$. At the next
moment an observer in $I$ sees that the three light rays arrive
simultaneously not at point $B$, but at $B^{\prime }$ since for the time $%
t=h/c$ the light rays travel toward $B$ the elevator moves at a
distance $\delta =at^{2}/2=ah^{2}/2c^{2}$. As the simultaneous
arrival of the three rays at point $B^{\prime }$ as viewed in $I$
is an absolute event being a {\em point} event, it follows that
the rays arrive simultaneously at $B^{\prime }$ as seen from
$N^{a}$ as well. Since for the {\em same} coordinate time $t=h/c$
in $N^{a}$ the three light rays travel different distances
$DB^{\prime }\approx h$, $AB^{\prime }=r+\delta $, and
$CB^{\prime }=r-\delta $ before arriving simultaneously at
point $B^{\prime }$ an observer in the elevator concludes that the {\em %
average} downward velocity $c_{\downarrow }^{a}$ of the light ray
propagating from $A$ to $B^{\prime }$ is slightly greater than $c$

\begin{equation}
c_{\downarrow }^{a}=\frac{h+\delta }{t}\approx c\left( 1+\frac{ah}{2c^{2}}%
\right) .  \label{c_a_dwn}
\end{equation}
The average upward velocity $c_{\uparrow }^{a}$ of the light ray propagating
from $C$ to $B^{\prime }$ is slightly smaller than $c$

\begin{equation}
c_{\uparrow }^{a}=\frac{h-\delta }{t}\approx c\left( 1-\frac{ah}{2c^{2}}%
\right) .  \label{c_a_up}
\end{equation}
Therefore the average velocities of light rays traveling parallel
and anti-parallel to the elevator's acceleration are different
from $c$\ and from each other. An observer in $N^{a}$ will
conclude that the three light rays arrive at $B^{\prime }$ (not
at $B$) due to the anisotropy in the propagation of light in the
elevator which in turn is caused by its accelerated motion.

\begin{center}
\begin {picture}(0,110)(-40,130)
\setlength{\unitlength}{0.25mm}
\put(-100,100){\framebox(110,220){}}
\put(10,212){\circle*{4}}
\put(-6,217){$B$}
\put(29,212){\line(1,0){10}}
\put(34,201){\vector(0,1){11}}
\put(10,190){\circle*{4}}
\put(-6,175){$B^{\prime}$}
\put(29,190){\line(1,0){10}}
\put(34,201){\vector(0,-1){11}}
\put(40,198){$\delta =\frac{ah^{2}}{2c^{2}}$}
\put(-5,305){$A$}
\put(10,320){\circle*{4}}
\put(10,100){\circle*{4}}
\put(-5,105){$C$}
\put(-100.5,212){\circle*{4}}
\put(-95,217){$D$}

\qbezier(-100.5,212)(-20,212)(2.5,193)
\put(2.5,193.3){\vector(3,-2){2}}
\put(-121,96.6){---}
\put(-121,317.1){---}
\put(-114,200){\vector(0,1){119.5}}
\put(-114,220.1){\vector(0,-1){119}}
\put(-131,212){$2h$}
\put(-100,80){\line(0,1){10}}
\put(10,80){\line(0,1){10}}
\put(-45,85){\vector(1,0){53.7}}
\put(-45,85){\vector(-1,0){53.7}}
\put(-50,87.5){$h$}
\put(17,320.5){\vector(0,-1){129}}
\put(17,99){\vector(0,1){89}}
\thicklines
\put(-45,255){\vector(0,1){33}}
\put(-39,267){{\bf a}}
\end {picture}
\end{center}

\vspace{2cm}

\begin{center}
\begin{list}{}{\leftmargin=1em \rightmargin=0em}\item[]
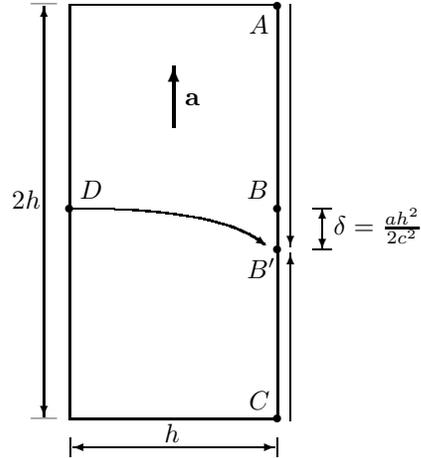
{\bf Figure 1}. Three light rays propagate in an accelerating
elevator. After having been emitted simultaneously from points
$A$, $C$, and $D$ the rays meet at $B^{\prime}$. The ray
propagating from $D$ toward $B$, but arriving at $B^{\prime}$,
represents the original thought experiment considered by
Einstein. The light rays emitted from $A$ and $C$ are introduced
in order to determine the expression for the average anisotropic
velocity of light in an accelerating frame of reference. It takes
the same time $t=h/c$ for the rays to travel the distances
$DB^{\prime} \approx h$, $AB^{\prime}=h+\delta$, and
$CB^{\prime}=h-\delta$. Therefore the average velocity of the
downward ray from $A$ to $B^{\prime}$ is $c^{a}_{\downarrow}=
(h+\delta)/t \approx c(1+ah/2c^{2})$; the average velocity of the
upward ray from $C$ to $B^{\prime}$ is $c^{a}_{\uparrow}=
(h-\delta)/t \approx c(1-ah/2c^{2})$.
\end{list}
\end{center}

As seen from (\ref{c_a_dwn}) and (\ref{c_a_up}) the average
anisotropic velocity of light in $N^{a}$ involves accelerations
and distances for which $ah/2c^{2}<1$. This restriction is always
satisfied since it is weaker than the one imposed by the
principle of equivalence which requires that only small regions
in a gravitational field where the field is uniform are
considered \cite{pe} (see also the paragraph immediately after
equation (\ref{c_g_up})).

Consider now an elevator (i.e. a non-inertial reference frame
$N^{g}$) at rest in the Earth's gravitational field. The elevator
will appear accelerating upward (with an acceleration $g=\left|
{\bf g}\right| $) with respect to a reference frame $I$ which is
at rest with respect to $N^{g}$, but starts to fall in the
gravitational field at the moment the light rays are emitted.
During the time the light rays emitted from the points $A$, $C$,
and $D$ travel toward $B$ the elevator will appear to move with
respect to $I$ at a distance $\delta =gt^{2}/2=gh^{2}/2c^{2}$ and
for this reason the three light rays will meet not at $B$ but at
$B^{\prime }$ situated bellow $B$ at a distance $\delta $.
Therefore an observer in $N^{g}$ also finds that the propagation
of light is anisotropic in the elevator. The average velocity of
the light ray traveling from $A$ to $B^{\prime }$ along ${\bf g}$
is

\begin{equation}
c_{\downarrow }^{g}=\frac{h+\delta }{t}\approx c\left(
1+\frac{gh}{2c^{2}}\right) . \label{c_g_dwn}
\end{equation}
The average velocity of the light ray propagating from $C$ to $B^{\prime }$
is slightly smaller than $c$ \cite{note}:

\begin{equation}
c_{\uparrow }^{g}=\frac{h-\delta }{t}\approx c\left(
1-\frac{gh}{2c^{2}}\right) . \label{c_g_up}
\end{equation}

One may deduce from (\ref{c_g_up}) that $gh/2c^{2}<1$ but this is
an apparent restriction. The average velocity of light between two
points separated by a greater distance in a strong gravitational
field is determined in terms of the explicit difference of the
gravitational potentials of the two points \cite{strongg}.

The average anisotropic velocities (\ref{c_g_dwn}) and
(\ref{c_g_up}) describe the propagation of light between {\em
two} points in a gravitational field, separated by a distance
$h,$ since $gh$ is the difference of the gravitational potential
of the two points. It should be specifically noted that those
velocities can be obtained from the expression for the velocity
of light in a gravitational field derived by Einstein in 1911
$c^{\prime }=c\left( 1+\Delta \Phi /c^{2}\right) $
\cite{einstein}, where $\Delta \Phi $ is the difference of the
gravitational potential of the two points. This fact demonstrates
that the 1911 Einstein velocity of light, which led him to a
wrong value of the deflection of light by the Sun, has been
prematurely abandoned. In his 1916 paper~\cite{einstein16} he
obtained the correct deflection angle by using the {\em
coordinate} velocity of light $c^{\prime }~=~c(1~-~2GM/Rc^{2}).$
However, as this velocity depends on the gravitational potential
of one point only it does not describe the propagation of light
between two points in a gravitational field; it is the 1911
expression for the velocity of light that provides the correct
description of this case.

One can also get the average velocities (\ref{c_g_dwn}) and (\ref{c_g_up})
directly from (\ref{c_a_dwn}) and (\ref{c_a_up}) by using the equivalence
principle and substituting $a=g$ in (\ref{c_a_dwn}) and (\ref{c_a_up}) \cite{pe2}.

The velocities (\ref{c_g_dwn}) and (\ref{c_g_up}) demonstrate
that there exists a directional dependence in the propagation of
light between two points in a gravitational field. This
anisotropy in the propagation of light is an overlooked corollary
of general relativity. Therefore an experiment for testing that
anisotropy constitutes another test of general relativity.

The purpose of this paper is to propose an experiment to test the
average anisotropic velocity of light in a gravitational field.
The experiment to be described bellow is based on another
experiment which was proposed by Stolakis \cite{stolakis} in 1986
with the intention to measure the one-way velocity of light. It
turned out that this could not be done but it was pointed out
that the experiment he proposed might be used for testing a
possible anisotropy of spacetime \cite{PetkovBJPS}. The
experiment can be described in the following way. Consider again
the Einstein elevator at rest in the Earth's gravitational field
(Figure 2). At point $B$ a light beam is split into two beams $1$
and $2$ which propagate vertically (with respect to the Earth's
surface). Ray $1$ travels the distance $h$ upward from $B$ to $A$
in a medium of index of refraction $n$; at $A$ it is reflected by
a mirror and its return path toward $B$ is in vacuum ($n=1$). Ray
$2$ also travels the same distance $h$ in a medium of refractive
index $n$ but downward from $B$ to $C$; its return path after
being reflected by a mirror at $C$ is in vacuum too. Upon their
arrival at point $B$ rays $1$ and $2$ interfere. If the average
velocity of light is anisotropic the interference pattern
produced by vertically propagating rays will differ from the
interference pattern of two horizontally traveling rays.

\begin{center}
\begin{picture}(100,100)(-30,-150)
\put(-30,-300){\framebox(10,100){}}
\put(-30,-180){\framebox(10,100){}}
\put(20,-190){\circle{8}}
\put(17,-193){\vector(-1,0){40}}
\put(17,-187){\vector(-1,0){40}}
\put(-25,-186){\vector(0,1){121}}
\put(-25,-194){\vector(0,-1){121}}
\put(-27,-315.5){\vector(-1,0){17}}
\put(-27,-65){\vector(-1,0){17}}
\put(-45.5,-315){\vector(0,1){124}}
\put(-30,-318){\line(2,1){10}}
\put(-20,-189.7){\line(-2,1){10}}
\put(-30,-195){\line(2,1){10}}
\put(-43,-318){\line(-2,1){10}}
\put(-30,-61.6){\line(2,-1){10}}
\put(-53,-66.6){\line(2,1){10}}
\put(-45.5,-65){\vector(0,-1){124}}
\put(-40.7,-57){A}
\put(-40.7,-329){C}
\put(-40.7,-194){B}
\put(28,-194){S}
\put(-8.1,-79.7){\line(1,0){10}}
\put(-8.1,-180.2){\line(1,0){10}}
\put(-8.1,-199.7){\line(1,0){10}}
\put(-8.1,-300){\line(1,0){10}}
\put(-3.3,-129.3){\vector(0,-1){50}}
\put(-3.3,-129.3){\vector(0,1){48.3}}
\put(0,-129.3){$h$}
\put(-3.3,-250){\vector(0,-1){49}}
\put(-3.3,-250){\vector(0,1){49}}
\put(0,-250){$h$}
\put(-23,-83){\line(3,4){15}}
\put(-6,-61){$n$}
\put(-23,-297){\line(3,-4){15}}
\put(-6,-324){$n$}
\put(-55,-129.3){$1$}
\put(-55,-250){$2$}
\put(58,-193){${\bf g}$}
\thicklines
\put(50,-175){\vector(0,-1){33}}
\end{picture}
\end{center}

\vspace{6cm}

\begin{center}
\begin{list}{}{\leftmargin=1em \rightmargin=0em}\item[]
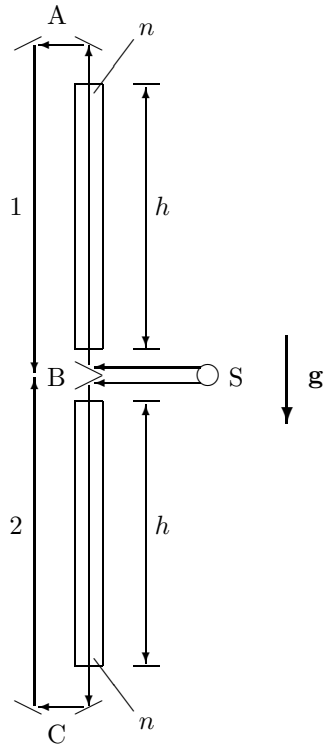
{\bf Figure 2}. An experiment for measuring the average
anisotropic velocity of light in a gravitational field of
strength ${\bf g}$.  At point $B$ a light ray coming from a
source $S$ is split into two rays $1$ and $2$ which propagate in
a medium of refractive index $n$ toward points $A$ and $C$,
respectively. After having traveled back to $B$ through a vacuum
the light rays interfere at $B$. Upon its arrival at $B$ light
ray $1$ (after having been reflected by a mirror at $A$) is
delayed by $\Delta t=(gh^{2}/c^{3})( n-1)$ with respect to ray
$2$ coming from $C$. The reason for the delay is that the
velocity of ray $1$ on its way from $B$ to $A$ is doubly slowed
down - by the medium and the gravitational filed.
\end{list}
\end{center}

Taking into account (\ref{c_g_up}) and (\ref{c_g_dwn}) the upward average
velocity of ray $1$ from $B$ to $A$ in the medium is

\[
c_{n}^{\uparrow }=\frac{c_{\uparrow }^{g}}{n}=\frac{c}{n}\left( 1-\frac{gh}{%
2c^{2}}\right) .
\]
while the downward average velocity of ray $2$ from $B$ to $C$ also in the
medium is

\[
c_{n}^{\downarrow }=\frac{c_{\downarrow }^{g}}{n}=\frac{c}{n}\left( 1+\frac{%
gh}{2c^{2}}\right) .
\]
Then the time for which ray $1$ goes to $A$ through the medium and returns
to $B$ through the vacuum is

\begin{equation}
t^{1}=\frac{h}{c_{n}^{\uparrow }}+\frac{h}{c_{\downarrow }^{g}}\approx \frac{%
hn}{c}\left( 1+\frac{gh}{2c^{2}}\right) +\frac{h}{c}\left( 1-\frac{gh}{2c^{2}%
}\right) .  \label{t1}
\end{equation}
The time for which ray $2$ travels through the medium to $C$ and returns to $%
B$ in the vacuum is

\begin{equation}
t^{2}=\frac{h}{c_{n}^{\downarrow }}+\frac{h}{c_{\uparrow }^{g}}\approx \frac{%
hn}{c}\left( 1-\frac{gh}{2c^{2}}\right) +\frac{h}{c}\left( 1+\frac{gh}{2c^{2}%
}\right) .  \label{t2}
\end{equation}
The difference between the two time intervals $t^{1}$ and $t^{2}$ is

\begin{equation}
\Delta t=t^{1}-t^{2}=\frac{gh^{2}}{c^{3}}\left( n-1\right) .  \label{delta_t}
\end{equation}

It is seen from (\ref{t1}) and (\ref{t2}) that $\Delta t\neq 0$ (for $n>1$)
only if the average velocity of light is anisotropic, i.e. only if it is
different from $c$. If the return paths of rays $1$ and $2$ are in a medium
of index of refraction $n^{\prime }$ then obviously (\ref{delta_t}) becomes

\[
\Delta t=\frac{gh^{2}}{c^{3}}\left( n-n^{\prime }\right) .
\]

The delay $\Delta t$ is best visualized by using a spacetime diagram as
shown in Figure 3.

\begin{center}
\begin {picture}(0,150)(-40,150)
\setlength{\unitlength}{0.25mm}
\put(-146.5,100){\line(0,1){290}}
\put(-63.5,100){\line(0,1){290}}
\put(19.5,100){\line(0,1){290}}
\put(-63.6,107){\circle*{4}}

\put(-63.5,383){\circle*{4}}
\put(-63.5,357){\circle*{4}}
\dottedline{2}(-63.5,107)(-146.5,287)
\dashline{11}(-63.5,107)(-146.5,247)
\dottedline{2}(-63.5,107)(19.5,287)
\dottedline{2}(-63.5,107)(19.5,287)
\dottedline{2}(19.5,287)(-63.5,370)
\dottedline{2}(-146.5,287)(-63.5,370)
\dashline{11}(-146.5,247)(-63.5,357)
\dashline{11}(-63.5,107)(19.5,327)
\dashline{11}(19.5,327)(-63.5,383)

\put(16,80){$A$}
\put(-67.5,80){$B$}
\put(-150,80){$C$}

\put(-102.5,380){---}
\put(-102.5,354){---}
\put(-101,366){$\Delta t$}
\put(-95,377){\vector(0,1){6}}
\put(-95,364){\vector(0,-1){6}}
\put(-100,297){$2$}
\put(-20,360){$1$}
\put(-20,248){$1$}
\put(-130,195){$2$}

\put(-146.5,400){\line(0,1){10}}
\put(-63.5,400){\line(0,1){10}}
\put(19.5,400){\line(0,1){10}}
\put(-105,405){\vector(1,0){40}}
\put(-105,405){\vector(-1,0){40}}
\put(-22,405){\vector(1,0){40}}
\put(-22,405){\vector(-1,0){40}}
\put(-106,410){$h$}
\put(-25,410){$h$}
\put(45,346){\vector(0,1){41}}
\put(49,364){$ct$}
\thicklines
\put(72,255){\vector(-1,0){40}}
\put(49,262){${\bf g}$}
\end {picture}
\end{center}

\vspace{3cm}

\begin{center}
\begin{list}{}{\leftmargin=1em \rightmargin=0em}\item[]
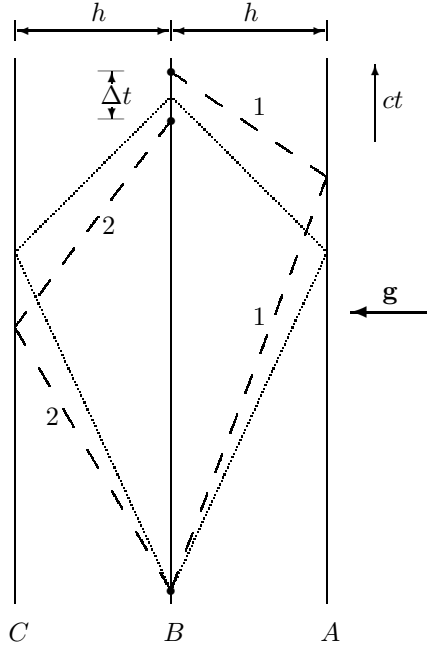
{\bf Figure 3}.  Spacetime diagram of the propagation of light
rays $1$ and $2$ as shown in Figure 2. If the average velocity of
light between two points in a gravitational field is anisotropic,
the worldlines of the rays will be represented by the dash lines;
if the velocity of light is $c$ the rays will be represented by
the dotted lines. The vertical worldlines depict points $A$, $B$,
and $C$.
\end{list}
\end{center}

Performed as described here the experiment cannot detect a change
in the interference pattern of two light rays one of which is
delayed by $\Delta t$ that is $\sim 10^{-23}s$ for $h=10m.$ This
tiny delay is equivalent to a shift of the wavelength of one of
the rays with respect to the other's wavelength by $10^{-5}nm$
which cannot produce an observable change in the interference
pattern. If, however, the two rays are made to pass multiple
times through the medium toward points $A$ and $C$, respectively
and through the vacuum back to point $B$, the delay $\Delta t$
will accumulate and the effect may become detectable \cite{manly}.


\begin{thebibliography}{99}
\bibitem{note}  The average velocities of light (\ref{c_g_dwn}) and (\ref
{c_g_up}) can be directly obtained by noting that the propagation of light
is affected by the Earth's gravity. At $B^{\prime }$ the instantaneous
velocity of the ray emitted at $C$ will be slowed down by the Earth's
gravity: $c_{\uparrow }^{B^{\prime }}=c-gt\approx c(1-gh/c^{2})$. As the
initial velocity of the ray is $c$ its average velocity from $C$ to $%
B^{\prime }$ is $c_{\uparrow }^{g}=c(1-gh/2c^{2})$. The ray
emitted at $A$ is falling in the Earth's gravitational filed
while traveling downward and will arrive at $B^{\prime }$ with
the instantaneous velocity $c_{\downarrow }^{B^{\prime
}}=c+gt\approx c(1+gh/c^{2})$ and the ray's average velocity from
$A$ to $B^{\prime }$ will be $c_{\downarrow }^{g}=c(1+gh/2c^{2})$.
\end{thebibliography}
\end{document}